\documentclass[useAMS,usenatbib]{mn2e}
\usepackage{graphicx}
\usepackage{epsfig}
\usepackage{epsf}
\usepackage{amsmath}
\usepackage{amssymb}
\usepackage{stfloats}
\usepackage{url}
\usepackage{longtable}
\usepackage{aas_macros}
\usepackage{deluxetable}

\newcommand\ion[2]{#1$\,${\small{#2}}\relax}

\title[SN\,2011fe]{Constraining the Progenitor Companion of the Nearby Type Ia SN\,2011fe with a Nebular Spectrum at +981 Days}
 
\author[Graham et al.]{M.~L. Graham$^{1}$\thanks{E-mail: melissagraham@berkeley.edu}, 
P.~E. Nugent$^{1,2}$, 
M. Sullivan$^{3}$, 
A.~V. Filippenko$^{1}$,
S.~B. Cenko$^{4,5}$,
\newauthor 
J.~M. Silverman$^{6}$,
K.~I. Clubb$^{1}$,
W. Zheng$^{1}$ \\
$^{1}$ Department of Astronomy, University of California, Berkeley, CA 94720-3411, USA \\
$^{2}$ Lawrence Berkeley National Laboratory, 1 Cyclotron Road, MS 90R4000, Berkeley, CA 94720, USA \\
$^{3}$ Department of Physics and Astronomy, University of Southampton, Southampton SO17 1BJ, United Kingdom \\
$^{4}$ Astrophysics Science Division, NASA Goddard Space Flight Center, MC 661, Greenbelt, MD 20771, USA \\
$^{5}$ Joint Space-Science Institute, University of Maryland, College Park, MD 20742, USA \\
$^{6}$ Department of Astronomy, University of Texas, Austin, TX 78712, USA 
}

\begin{document}
\pagerange{\pageref{firstpage}--\pageref{lastpage}} \pubyear{2014}

\maketitle

\label{firstpage}

\begin{abstract}

We present an optical nebular spectrum of the nearby Type Ia supernova 2011fe, obtained 981 days after explosion. SN\,2011fe exhibits little evolution since the +593 day optical spectrum, but there are several curious aspects in this new extremely late-time regime. We suggest that the persistence of the $\sim5800$~\AA\ feature is due to \ion{Na}{I}~D, and that a new emission feature at $\sim7300$~\AA\ may be [\ion{Ca}{II}]. Also, we discuss whether the new emission feature at $\sim6400$~\AA\ might be [\ion{Fe}{I}] or the high-velocity hydrogen predicted by Mazzali et al. The nebular feature at 5200~\AA\ exhibits a linear velocity evolution of $\sim350$ $\rm km\ s^{-1}$ per 100 days from at least +220 to +980 days, but the line's shape also changes in this time, suggesting that line blending contributes to the evolution. At $\sim 1000$ days after explosion, flux from the SN has declined to a point where contribution from a luminous secondary could be detected. In this work we make the first observational tests for a post-impact remnant star and constrain its temperature and luminosity to $T \gtrsim 10^4$ $\rm K$ and $L \lesssim 10^4$ $\rm L_{\odot}$. Additionally, we do not see any evidence for narrow H$\alpha$ emission in our spectrum. We conclude that observations continue to strongly exclude many single-degenerate scenarios for SN\,2011fe.

\end{abstract}


\section{Introduction} \label{sec:intro}

Supernovae of Type Ia (SNe\,Ia) are thermonuclear explosions of carbon-oxygen white-dwarf stars. As standardisable candles they are a very valuable tool for modern cosmology, but their progenitor scenario(s) and explosion mechanism(s) are not yet well understood (e.g., Howell 2011). For the former, the two leading  candidates are the double-degenerate scenario, in which the primary merges with another carbon-oxygen white dwarf or accretes matter from a helium white dwarf, and the single-degenerate scenario, in which matter is accreted from a red-giant or main-sequence companion. Most of the current efforts toward identifying the progenitor are focused on pre- or post-explosion imaging to directly detect the progenitor star and/or its companion (e.g., Li et al. 2011; Kelly et al. 2014), and looking for evidence of the SN\,Ia ejecta interacting with circumstellar material (CSM) --- the amount, composition, and distribution of which theoretically depends on the progenitor system (e.g., Leonard 2007; Patat et al. 2007; Sternberg et al. 2014). 

SN\,2011fe was discovered by the Palomar Transient Factory (Rau et al. 2009; Law et al. 2009) within hours of its explosion on 23.7 Aug. 2011 (dates are UT throughout this paper) and spectroscopically classified as a normal SN\,Ia (Nugent et al. 2011). The host galaxy is M101 (NGC 5457), just $6.4\pm0.7$ Mpc away (Shappee \& Stanek 2011) with a recession velocity of $v_{\rm{rec}} = 241$ $\rm km\ s^{-1}$ (de Vaucouleurs et al. 1991). Richmond \& Smith (2012) present photometric monitoring, reporting that SN\,2011fe reached peak brightness on 11.5 Sep. 2011. Nugent et al. (2011) used the very early-time light curve to constrain the size of the progenitor to white-dwarf scale (later refined by Bloom et al. 2012), and they show that SN\,2011fe exhibited no evidence of interaction between the SN ejecta and the companion star as predicted by Kasen (2010). Li et al. (2011) use pre-explosion {\it Hubble Space Telescope (HST)} archival images to exclude bright red giants and most main-sequence helium stars as mass-donor companions to the white dwarf. Graur et al. (2014) show that archival narrow-band {\it HST} images centred on \ion{He}{II} $\lambda$4686 indicate no supersoft X-ray source existed at this location, contrary to expectations for the single-degenerate scenario. These studies eliminate many single-degenerate companions for SN\,2011fe.

The proximity of SN\,2011fe allows it to be monitored to extremely late times (e.g., Kerzendorf et al. 2014; Taubenberger et al. 2014), and in this evolved state two tests of the progenitor system become feasible. First, the lower luminosity of the SN\,Ia may allow for even a relatively small mass of hydrogen in the system to be observed via H$\alpha$ emission (e.g., Marietta et al. 2000; Leonard \& Filippenko 2001; Mattila et al. 2005; Leonard 2007; Lundqvist et al. 2013; Shappee et al. 2013a; Lundqvist et al. 2015). Hydrogen is expected to be present if the companion was a red giant or main-sequence star, but not in any appreciable amount if the companion was a carbon-oxygen or helium white dwarf. Second, the lower luminosity of the SN\,Ia may allow for a direct detection of the remaining companion star in the single-degenerate scenario, because the secondary is theoretically predicted to increase in temperature and luminosity in the years after impact with the SN\,Ia ejecta (e.g., Pan et al. 2013; Liu et al. 2013; Shappee et al. 2013b).

In this work, we use an optical nebular spectrum obtained 981 days after explosion to constrain the amount of hydrogen present in the system, and the temperature and luminosity of a putative post-impact remnant companion star. In \S \ref{sec:obs} we present our observations, and in \S \ref{sec:ana} we analyse our spectrum, comparing it to previous nebular observations in 2013 and 2012. Section \ref{ssec:anaH} demonstrates a nondetection of H$\alpha$ and places an upper limit on the mass of hydrogen in the system, and \S \ref{ssec:anacomp} provides two independent tests for the signature of a post-impact remnant star (PIRS) in our spectrum of SN\,2011fe. We summarise and conclude in \S \ref{sec:con}.

\section{Observations}\label{sec:obs}

We obtained spectra of SN\,2011fe with the Low Resolution Imaging Spectrometer (LRIS; Oke et al. 1995) at the Keck Observatory on 2014 May 01 --- that is, 981 days after explosion (we use observer-frame time). At such a late phase, the source was too faint to appear in the LRIS guide-camera images recorded during spectroscopy, so $\sim3.5$ weeks later we obtained images in the $g$ and $R_S$ bands with LRIS on 2014 May 25 (+1006 days). Here we present and discuss first the imaging in \S~\ref{ssec:obsimage} and then the spectroscopy in \S~\ref{ssec:obsspec}. All of the spectra are available on WISeREP\footnote{http://wiserep.weizmann.ac.il} (Yaron \& Gal-Yam 2012).

\subsection{Imaging}\label{ssec:obsimage}

\begin{figure}
\begin{center}
\includegraphics[trim=0.0cm 1.5cm 0.0cm 0cm,clip=true,width=3.3in]{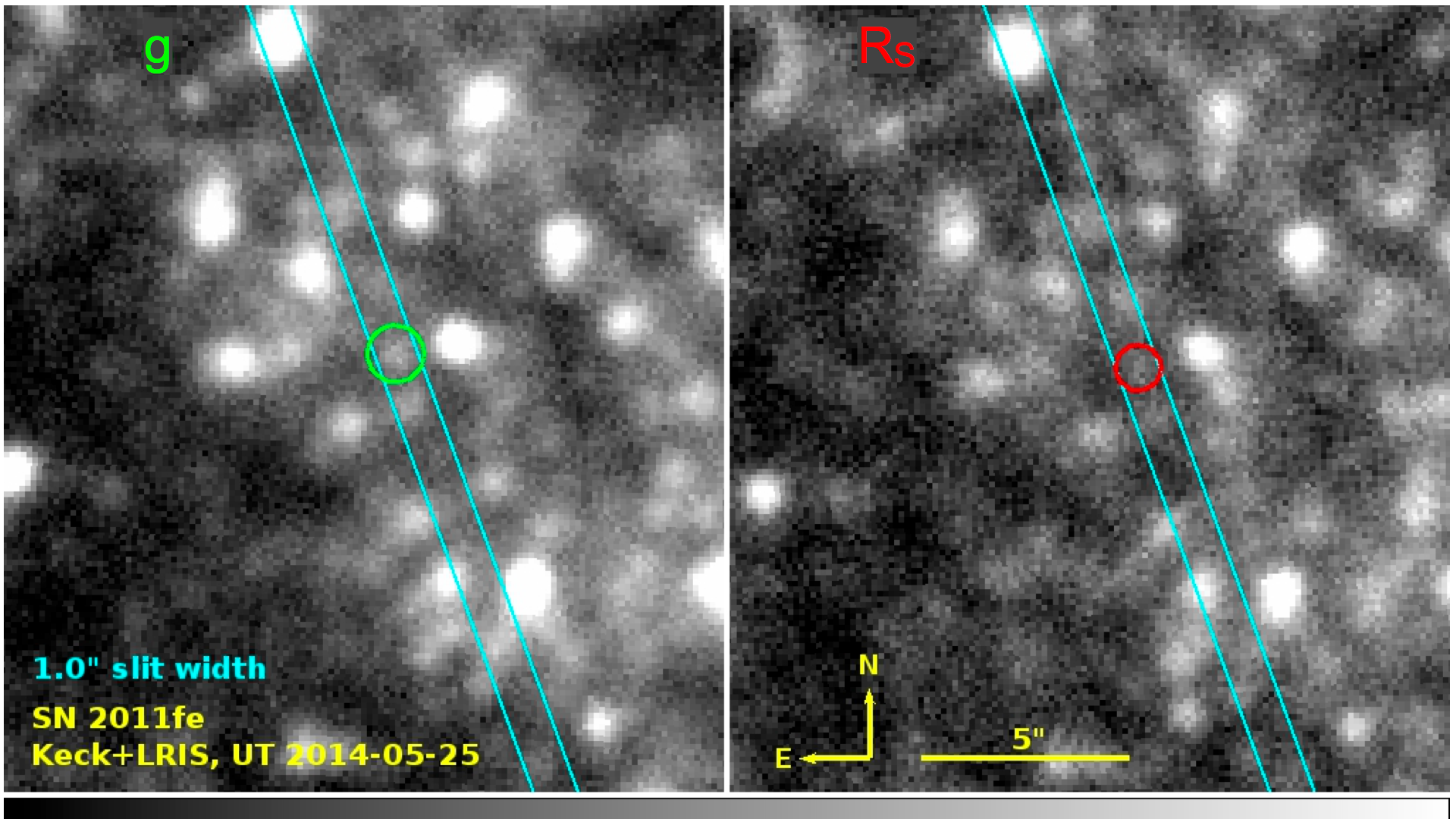}
\caption{Images from Keck+LRIS on 2014 May 25 in the LRIS $g$ and $R_S$ filters (left and right, respectively). Pixels have been grey-scaled to show the faint  SN\,2011fe (circle) in a field crowded by brighter sources and contaminated by diffuse background light from M101. The blue box represents the slit width and orientation used for the Keck+LRIS spectrum obtained on 2014 May 1, and the size of the circles is set equal to the full width of the aperture used to extract the 1-dimensional spectrum. Despite the crowded field, our slit width, orientation, and chosen aperture limit contamination from nearby stars. \label{fig:img}}
\end{center}
\end{figure}

With Keck+LRIS, we acquired three images in the $g$ filter of 180, 220, and 230\,s duration (increasing the exposure time because the red-side readout time is significantly longer than that of the blue side), and three images in the $R_S$ filter of 180\,s each. SN\,2011fe is in a crowded field with much host-galaxy background light, and Sloan Digital Sky Survey (SDSS) standard stars are not available within our small field of view. For photometric calibration, we also obtained two 1\,s images in each of two regions in the CFHT Legacy Survey Deep Field 3 immediately after SN\,2011fe; these fields were chosen because they are well covered (e.g., by SDSS) and are only $\sim2^\circ$ from SN\,2011fe. We applied typical image reduction and analysis procedures to all fields, and used {\sc Source Extractor} (Bertin \& Arnouts 1996) to find and measure source photometry (using aggressive deblending parameters for the crowded field of SN\,2011fe). Our final stacked images are shown in Figure \ref{fig:img}.

On 2014 May 25, we had an LRIS blue-side shutter failure and the chip remained illuminated during readout. The blue-side readout time is 42\,s, so this affects the Deep Field images used to determine our zeropoint more than it affects the SN\,2011fe images. We use SDSS star magnitudes from Data Release 10 (Ahn et al. 2014) to calculate a simple zeropoint ($z_{g} = m_{g,{\rm SDSS}} - m_{g,{\rm raw}}$) and find that this increases by $\sim0.2$ mag across the chip. We take this into account by using the mean zeropoint, which also represents the zeropoint at the centre of the chip where SN\,2011fe is, and adding 0.2 in quadrature to the uncertainty. We estimate that the brightness at the position of SN\,2011fe was $m_g \approx 24.3\pm0.3$ mag on 2014 May 25. The line-of-sight Galactic extinction is much smaller than our uncertainty, $A_V=0.023$\, mag from the NASA Extragalactic Database (NED\footnote{http://ned.ipac.caltech.edu/}; Schlafly \& Finkbeiner 2011; Schlegel, Finkbeiner \& Davis 1998). For comparison, Kerzendorf et al. (2014) find that $m_g = 23.43 \pm 0.28$\,mag on 2014 March 7; assuming a decline rate of 0.01\,mag per day, the predicted brightness for our observation is $m_g \approx 24.0$\,mag. Given our large uncertainties, our results are not discrepant with this prediction --- however, there is evidence that the rate of decline is probably slower than 0.01\,mag per day at this very late phase (e.g., Kerzendorf et al. 2014). On the red side, we estimate $m_R \approx24.6$\,mag with similarly large uncertainties. As we do not use this in our analysis, we leave a more accurate photometric measurement for a future publication, including template image subtraction to remove background light. Note that Kerzendorf et al. (2014) also have not subtracted a template image.

In \S \ref{ssec:anacomp}, our constraints on the presence of a PIRS depend on an accurate flux calibration for our spectrum, especially at the blue end, and our results are more conservative if we adopt a brighter $g$-band magnitude. Synthesised spectrophotometry from our reduced Keck+LRIS spectrum (described below) indicates $m_g \approx 23.9 \pm 0.2$\,mag and $R_S \gtrsim 24.2 \pm 0.5$\,mag, in agreement with (but slightly brighter than) our photometry. Since this is quite sufficient for the analysis, we make no further adjustments to the flux calibration of the spectrum.

\subsection{Spectroscopy}\label{ssec:obsspec}

For our LRIS spectra we used the 1.0\arcsec\ slit rotated to the parallactic angle (Filippenko 1982), which was $200^\circ$ east of north, to minimise the effects of atmospheric dispersion (in addition, LRIS has an atmospheric dispersion corrector). The size and orientation of the slit is shown as the blue box in Figure \ref{fig:img}. An offset star was used for target acquisition because SN\,2011fe was too faint to appear in the guider camera. In our LRIS configuration, coverage in the blue with the 600/4000 grism extends over $\lambda=$ 3010--5600~\AA, with a dispersion of 0.63~\AA\ pixel$^{-1}$ and a full width at half-maximum intensity (FWHM) resolution of $\sim 4$~\AA. Coverage in the red with the 400/8500 grating extends over $\lambda=$ 4500--9000~\AA, with a dispersion of 1.16~\AA\ pixel$^{-1}$ and a resolution of FWHM $\approx 7$~\AA. We did not dither along the slit between exposures, instead staying at the same CCD location in all four 1200~s exposures.

These spectra were reduced using routines written specifically for LRIS in the Carnegie {\sc Python} ({\sc CarPy}) package. The two-dimensional (2D) images were flat fielded, corrected for distortion along the $y$ (slit) axis (i.e., rectified), wavelength calibrated, cleaned of cosmic rays (imperfect), and stacked with a median combine before defining the apertures and extracting the one-dimensional (1D) spectra (sky subtraction is discussed below). The width of the extraction aperture was $\pm5$ pixels ($\pm0.68$\arcsec) on the blue side and $\pm4$ pixels ($\pm0.54$\arcsec) on the red side, as shown by the radius of the green and red circles in Figure \ref{fig:img}. The seeing is estimated to have been $\sim0.9$\arcsec (FWHM) during the observations. Although the field is crowded, Figure \ref{fig:img} shows that the nearest neighbours do not contaminate our spectrum. Blue and red standard stars were used to determine the sensitivity functions and flux calibrate the data for each side, respectively.

\begin{figure}
\begin{center}
\includegraphics[trim=1cm 0.2cm 0.0cm 0.5cm,clip=true,width=3.4in]{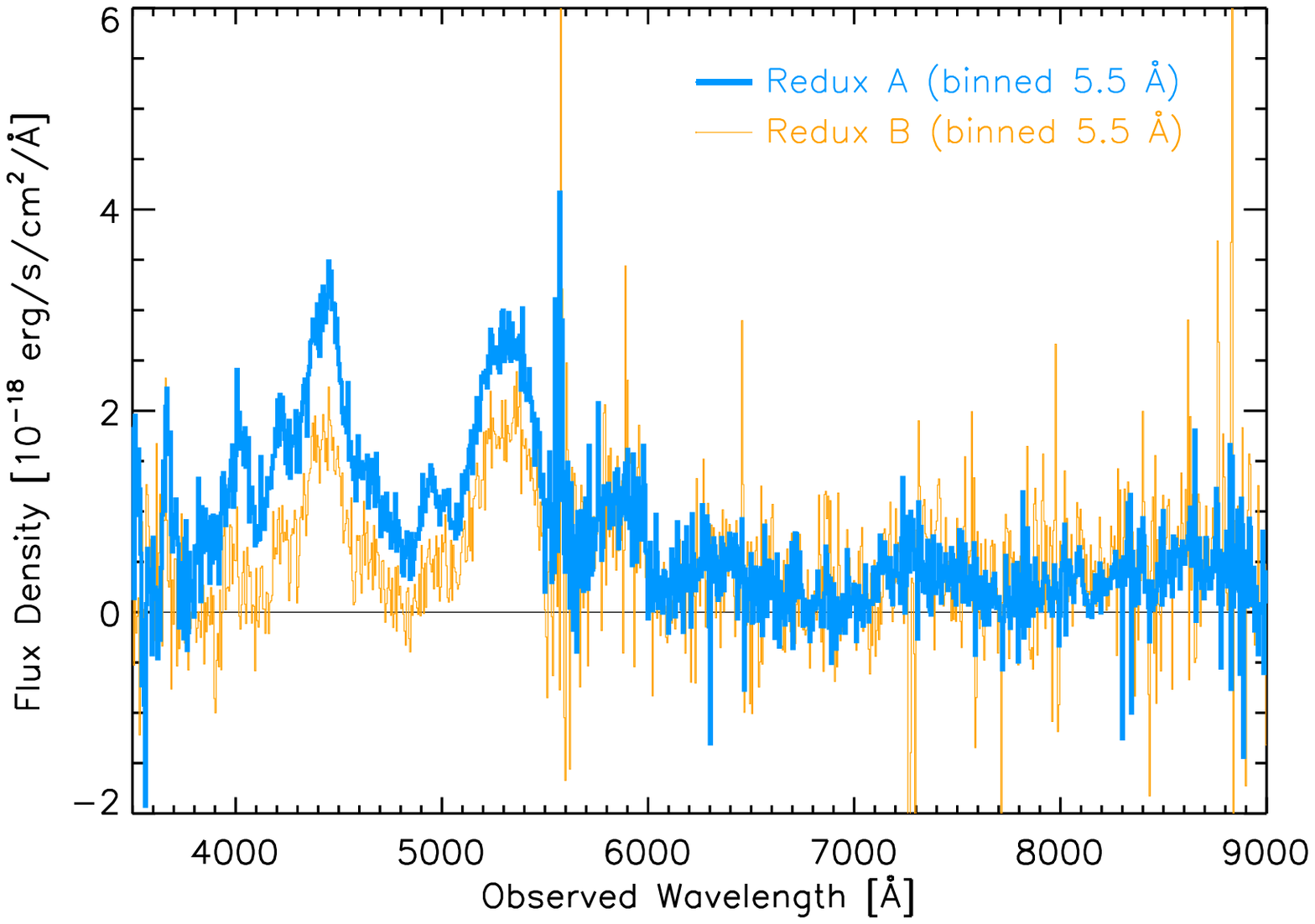}
\includegraphics[trim=1cm 0.2cm 0.0cm 0.5cm,clip=true,width=3.4in]{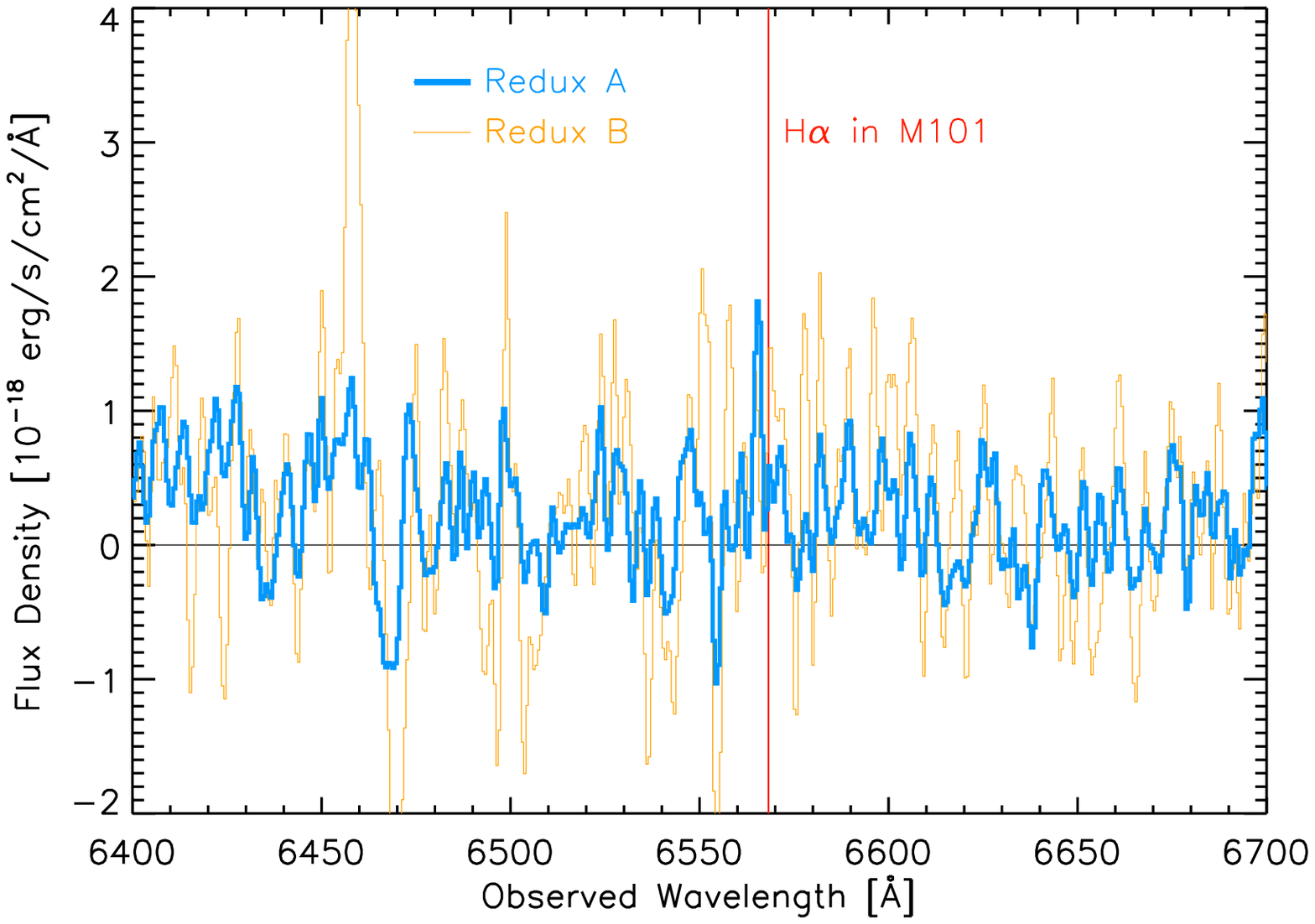}
\caption{The reduced and calibrated LRIS spectra of SN\,2011fe, showing the small systematic difference introduced by the treatment of the sky background for this faint object. For ``Redux A" (blue) we created a sky frame for each 2D image and subtracted it prior to the median combine and aperture extraction; for ``Redux B" (orange) we defined background apertures on either side of the trace during aperture extraction from the median-combined 2D image. We show the resulting spectra for the whole wavelength range (top) and the region around H$\alpha$ only (bottom), in which the solid red line marks the expected position of H$\alpha$ at the recession velocity of SN\,2011fe. Spectra are binned as described on the plot; horizontal lines at zero flux are included to guide the eye. The excess noise at 5600~\AA\ is at the dichroic's crossover wavelength. \label{fig:comp5}}
\end{center}
\end{figure}

The input parameter space for these routines was carefully explored to ensure the best possible results, and we found that the method for removing the sky background had the largest influence on the final spectrum. The two options for treating the sky background are as follows. \textbf{``Redux A'':} create a sky frame for each 2D image and subtract it prior to median combining the images and performing aperture extraction (Kelson 2003). \textbf{``Redux B'':} define background apertures on either side of the trace during aperture extraction from the stacked 2D image and subtract from the 1D spectrum (i.e., set the background parameter to ``fit'' in {\sc iraf} task \textit{apall}). We used 4-pixel apertures adjacent to the extraction aperture. The results of these two reduction versions are compared in Figure \ref{fig:comp5}. 

Sky subtraction is the largest (although still small) source of systematic uncertainty. On the blue side ($\lambda\lesssim5500$~\AA), the continuum of ``Redux A" does not go to zero flux at $\lambda\approx4900$~\AA\ like the continuum of ``Redux B." This is not necessary, as both residual photospheric emission and low-level flux from emission lines can produce some continuum flux which is actually from the SN. However, it might indicate that an incomplete sky subtraction is achieved on the blue side when a sky frame for each 2D image is created and subtracted. In Figure \ref{fig:comp5} we can see that the choice of sky subtraction leads to a small difference in relative emission-line strengths on the blue side, but this will not affect our particular analysis. On the red side, the sky background is concentrated more into emission lines, and has less flux in the continuum, than on the blue side. On the red side, however, ``Redux A" has a lower scatter and fewer residual spikes from the sky emission lines because the 2D subtraction is more sophisticated than the 1D background apertures used in ``Redux B." We proceed using both versions of our reductions so that our conclusions are robust to systematics introduced by the data reduction for this very faint SN. Where necessary, the following analysis will state which reduction version is being used. Before any analysis is performed, the spectra are deredshifted into the rest frame of the host galaxy, M101, and corrected for line-of-sight Galactic extinction $E(B-V) \approx 0.008$\,mag (Schlegel, Finkbeiner, \& Davis 1998).

\subsection{Earlier Nebular Spectra of SN\,2011fe}\label{ssec:obscomp}

\begin{figure}
\begin{center}
\includegraphics[trim=1cm 0.2cm 0.0cm 0.5cm,clip=true,width=3.4in]{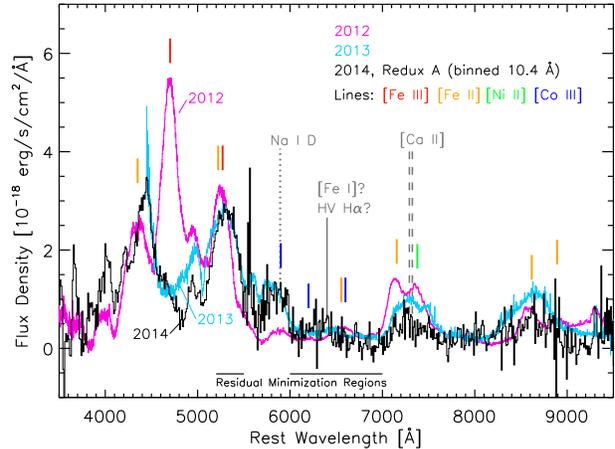}
\caption{A comparison of our spectrum of SN\,2011fe from 2014 (black) to nebular spectra from 2012 (pink) and 2013 (blue). The latter two have been flux scaled to the 2014 spectrum by determining the normalisation factor that minimises the absolute value of their flux residuals in the wavelength ranges specified on the plot ($0.005$ and $0.122$ for the 2012 and 2013 spectra respectively, as described in the text of \S~\ref{ssec:obscomp}; no wavelength shifts are applied here). The ordinate values therefore apply only to the 2014 spectrum. The ionised elements commonly attributed to each line are labeled with coloured bars, and our proposed line identifications are labeled in grey. While the new 2014 spectrum resembles the 2013 spectrum, both are significantly different from the 2012 spectrum, particularly in the lines of [\ion{Fe}{III}]. \label{fig:compspec}}
\end{center}
\end{figure}

For comparison, we also use nebular spectra of SN\,2011fe obtained under the Berkeley Supernova Ia Program (BSNIP; Silverman et al. 2012) in 2012 and 2013, shown in Figure \ref{fig:compspec}. The 2012 spectrum was taken on 23 Aug. 2012, 365 days after explosion, with the Kast spectrograph at Lick Observatory (Miller \& Stone 1993), using the 600/4310 grism and the 300/7500 grating, a long slit 2.0\arcsec\ wide, and a 1800~s exposure time. This spectrum is previously unpublished and will also appear in Mazzali et al. (2015). The 2013 spectrum was obtained on 8 April 2013, 593 days after explosion, with DEIMOS (Faber et al. 2003) on Keck II, using the 600 lines mm$^{-1}$ grating, the GG455 order-blocking filter, the 1.0\arcsec\  slit of the long variable multislit (LVM) slitmask, and a 1200~s exposure time. This spectrum is previously unpublished. The data were reduced using common reduction and analysis procedures (e.g., Silverman et al. 2012). 

In Figure \ref{fig:compspec} we have flux scaled the 2012 and 2013 spectra to the 2014 spectrum by multiplying the 2012 and 2013 spectral flux by constant normalisation factors. The normalisation factors are set to values that minimise the mean absolute flux residuals in particular wavelength regions ($0.005$ and $0.122$ for the 2012 and 2013 spectra respectively). These regions, shown in Figure \ref{fig:compspec}, are chosen to be areas on the red side in which the feature shapes were well matched between epochs and minimal evolution is expected. The red side was also chosen for this minimisation because most of the analysis will be done on the blue side. 

In \S \ref{sssec:anavar} we want to compare the variance of the 2014 spectrum with that from 2013, but unfortunately the DEIMOS spectrum from 2013 does not extend bluer than $\sim4500$~\AA. As a compromise, since the region blueward of 4500~\AA\ exhibits similar features in 2012 and 2014 (Figure \ref{fig:compspec}), we supplement the 2013 spectrum with the 2012 spectrum in the range 3500--4500~\AA\ to create a ``composite" 2012+2013 SN\,2011fe spectrum. First, the 2012 spectrum is shifted in wavelength such that the broad peaks at 4400 and 5200~\AA\ match with their positions in 2014 ($\sim60$~\AA). Then, the 2012 and 2013 spectra are flux scaled to the 2014 spectrum as described above, and then joined to produce a final composite spectrum. The application of this composite spectrum is described in \S \ref{sssec:anavar}.

\section{Analysis}\label{sec:ana}

Early-time photospheric-phase optical spectra of SN\,2011fe were analysed by Parrent et al. (2012), who classify it as spectroscopically normal and a member of the ``low velocity gradient" group, referring to the change in velocity of the \ion{Si}{II} $\lambda$6355 line (Benetti et al. 2005). Mazzali et al. (2014) apply abundance tomography models to near-ultraviolet and optical spectra of SN\,2011fe around maximum light and show that $^{56}$Ni is distributed in an extended region; that SN\,2011fe probably had a lower density and larger volume of nucleosynthetic products; and that the amount of iron in the outer layers of SN\,2011fe is consistent with subsolar metallicity, $Z_{\rm 11fe}\approx 0.5\,{\rm Z}_{\odot}$. A subsolar metallicity is also supported by the comparative analysis of photospheric through nebular-phase spectra of SN\,2011fe and SN\,2011by by Graham et al. (2015). The extensive collection of analyses of SN\,2011fe agree that it was a typical example of a SN~Ia.

In Figure \ref{fig:compspec} we compare our 2014 spectrum with flux-scaled spectra of SN\,2011fe from 2013 and 2012. We find that over the last year there has been little evolution in the spectral features; the 2013 and 2014 spectra are very similar to each other, and different from the 2012 nebular spectrum. The similarity of the spectral features as the total flux declines is evidence of no (or negligible) contributions from a light echo, consistent with previous findings that SN\,2011fe exploded in a ``clean'' environment (e.g., Chomiuk et al. 2012; Horesh et al. 2012; Patat et al. 2013; Pereira et al. 2013). We point out the following changes between the 2012 and 2014 spectra:
(1) the decline of the [\ion{Fe}{III}] line at $\lambda4700$~\AA,
(2) the feature at $\sim5800$~\AA\ where, at earlier times, [\ion{Co}{III}] appears,
(3) the potential rise of a new broad feature at 6400~\AA, just blueward of where the [\ion{Co}{III}] $\sim6600$~\AA\ feature declined after 2012,
(4) a decline in the [\ion{Fe}{II}] and [\ion{Ni}{II}] feature at $\sim7300$~\AA, and
(5) a general redshifting of all spectral features. 
These changes are discussed below (and in some cases also by Taubenberger et al. 2014).

The decline of doubly ionised species [\ion{Fe}{III}] and [\ion{Co}{III}] can be attributed to the decreasing temperature in the remnant ejecta material. Curiously, the feature at 5800~\AA, commonly thought to be [\ion{Co}{III}], appears quite strongly in both our 2013 and 2014 spectra, but is not seen by Taubenberger et al. (2014). We think this discrepancy is caused by a coincidence of this feature with their spectrograph's dichroic at 5750~\AA, as we cannot attribute this feature to be an artifact of our reduction pipeline (it appears quite plainly in our raw, sensitivity- and flux-calibrated data, for all versions of our reduction process). Given the decline of the [\ion{Co}{III}] 6600~\AA\ feature, another species is probably responsible. [\ion{Fe}{I}] is a likely culprit, and it has plenty of transitions in this region, such as 5640, 5695, and 5709~\AA\ (Smith \& Wiese 1973). However, this line could also be \ion{Na}{I}~D, which has long been predicted at this location by models, and first considered (but rejected) in nebular spectra by Kuchner et al. (1994). Dessart et al. (2014) present a review of the literature for and against \ion{Na}{I}~D at $>100$ days (see their Section 2), and show that [\ion{Co}{III}] is the main contributor for the first two months after peak brightness, but at later times the Na may finally be visible because it does not decay like $^{56}$Co. For example, \ion{Na}{I}~D has been attributed to the 5800~\AA\ feature for the normal SN\,Ia 2003du by Tanaka et al. (2011). 

Regarding the rise of the broad feature at 6400~\AA, Taubenberger et al. (2014) attribute this to [\ion{Fe}{I}] or [\ion{O}{I}], pointing out that detailed spectral modeling is required to confirm it. We discuss in \S~\ref{sssec:anaHbroad} the possibility that this is broad, high-velocity emission from hydrogen. Further redward, we see that the neighbouring [\ion{Fe}{II}] $\lambda$7155 and [\ion{Ni}{II}] $\lambda$7370 features have declined, and that this region is now dominated by a third peak at $\sim7300$~\AA, which could be the redshifted line of [\ion{Fe}{II}], or [\ion{Ca}{II}] $\lambda\lambda$7292, 7324. Emission from Na and Ca may suggest the presence of circumstellar material from the progenitor system. 

We find that the continuous shift in the nebular features toward redder wavelengths is part of a linear trend that starts by at least 200 days after peak brightness. We measure the central wavelength of the [\ion{Fe}{II}] 5200~\AA\ feature in the 2014, 2013, and 2012 nebular spectra used in this work, as well as in the +300 and +200 day spectra from Graham et al. (2015), by smoothing the spectra and finding the wavelength at the centre of the FWHM. We find a linear evolution of $\sim6$~\AA\ ($\sim350$ $\rm km\ s^{-1}$) per 100 days for SN\,2011fe. This is somewhat lower than the typical evolution found by Silverman et al. (2013) for a larger sample of SN\,Ia nebular spectra\footnote{Note that Silverman et al. (2013) use the velocity of the [\ion{Fe}{III}] $\sim4700$~\AA\ feature, whereas we must use the [\ion{Fe}{II}] $\sim5200$~\AA\ feature, because the former does not appear in our 2013 or 2014 spectra.}, for which the average velocity of the nebular features decreases by 1000--2000 $\rm km\ s^{-1}$ per 100 days. They also find that the velocity appears to evolve to $v\approx0$ $\rm km\ s^{-1}$ by $\sim300$ days past maximum brightness for the SNe\,Ia in their sample (e.g., see their Figure 4), suggesting that continued velocity evolution until +1000 days results in a redshift. Taubenberger et al. (2014) also interpret their 2014 spectrum of SN\,2011fe as exhibiting a redshift in the rest frame. However, for SN\,2011fe, the 5200~\AA\ feature is not symmetric, and we see that the wavelength of the peak flux moves from the blue to the red side of the feature over time. Instead of using the wavelength at the centre of the FWHM, if we use the wavelength at \textit{peak} flux to calculate the linear evolution in the line velocity, we find $\sim11$~\AA\ ($\sim640$ $\rm km\ s^{-1}$) per 100 days for SN\,2011fe (i.e., the line's peak moves more rapidly redward than the line's centre). The fact that the line morphology changes significantly suggests that line blending cannot be ruled out as the source of the apparent shift, and we conclude that we cannot be sure that the nebular features are exhibiting a redshift in the rest frame. We hope that future spectral modeling and/or observations of extremely late-time SNe\,Ia will provide explanations for all of these observations.

In 2014, the flux of SN\,2011fe has declined far enough that tests of the progenitor companion using a nebular-phase spectrum are possible. In the following sections we perform three tests to constrain the properties of a putative companion star.

\subsection{An Upper Limit on the Mass of Hydrogen in the Progenitor System of SN\,2011fe}\label{ssec:anaH}

Type I supernovae are classified by a lack of hydrogen in their spectrum, but some progenitor scenarios, such as mass accretion from a main-sequence or red-giant star, predict that some hydrogen can be seen. For example, hydrogen could be stripped from a main-sequence star and be moving with $\lesssim 1000$ km s$^{-1}$ ($d\lambda \approx 22$~\AA; Liu et al. 2013, Figure 9). Nugent et al. (2011) found that there was no main-sequence or red-giant companion along our line of sight to SN\,2011fe, as the interaction between SN ejecta and such a star would increase the blue flux at very early times (Kasen 2010), but a main-sequence star might have existed on the far side. If hydrogen is present in the system, we might be able to see it at late times -- and if we do, then hydrogen cannot be ruled out as being absent in any SN\,Ia that has only early-time nondetections of hydrogen. We have carefully and thoroughly scrutinised our 2D reduced spectra but find no trace of an emission line at or anywhere near the position of SN\,2011fe, before or after background subtraction. (We do see host-galaxy emission in the lower half of the LRIS detector, on the other side of the chip gap from SN\,2011fe.)

As shown in the bottom plot of Figure \ref{fig:comp5}, there is no obvious emission seen in the 1D spectrum either. To quantify the maximum potential equivalent width of an H$\alpha$ emission line ``hiding" in our data, we follow a similar analysis to that of Shappee et al. (2013a), who constrain the amount of hydrogen in the system of SN\,2011fe using their deep spectra taken up to 275 days past maximum light.

Marietta et al. (2000) model the evolution of hydrogen from the companion star to 380 days after maximum brightness. Applying this model to SN\,Ia 2001el (which was normal, like SN\,2011fe), Mattila et al. (2005) find that $M=0.05~{\rm M}_{\odot}$ of material would create an H$\alpha$ line with peak luminosity $\sim 3.36 \times 10^{35}$ $\rm erg\ s\ \AA^{-1}$. Shappee et al. (2013a) find that for the distance and Galactic reddening of SN\,2011fe, this would be an observed peak of $6.71 \times 10^{-17}$ $\rm erg\ s^{-1}\ cm^{-2}\ \AA^{-1}$. An expression for the maximum equivalent width of an emission line ``hiding" in a spectrum, $W_{\lambda}(3\sigma)$, is derived and presented by Leonard \& Filippenko (2001) and Leonard (2007): 

\begin{equation}
W_{\lambda} (3\sigma) = 3\ \Delta I \sqrt{W_{\rm line}\ \Delta X},
\end{equation}

\noindent
where $\Delta I$ is the 1$\sigma$ root-mean square (RMS) of the fluctuations around the normalised continuum, $W_{\rm line}=22$~\AA, and $\Delta X = 0.61$~\AA\ (the resolution or bin size). To measure $\Delta I$, we apply a least-squares polynomial smoothing filter (Savitzky \& Golay 1964) with a width of 163 pixels (100~\AA), subtract the data from the smoothed continuum, and find the square root of the mean of the squared residuals in the range 6480--6700~\AA. The value of $\Delta I$ is then this RMS fluctuation divided by the continuum. 

\begin{figure}
\begin{center}
\includegraphics[trim=1cm 0.2cm 0.0cm 0.5cm,clip=true,width=3.4in]{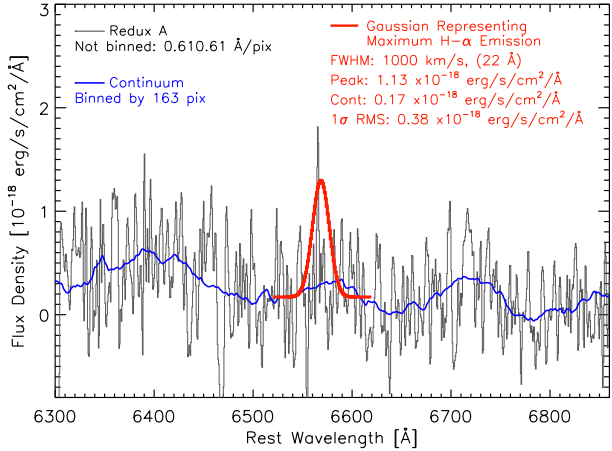}
\caption{The 1D spectrum from our reduction version ``A" (see \S \ref{sec:obs} and Figure \ref{fig:comp5}), shown in the rest frame of SN\,2011fe in the region of H$\alpha$. Our LRIS spectrum from 2014 is given in black, the continuum from polynomial smooth in blue, and a Gaussian profile representing the maximum possible H$\alpha$ feature in red. The Gaussian has a width of 22~\AA\ and height equal to 3 times the 1$\sigma$ RMS fluctuation around the continuum. (Note that the suspicious narrow line has a FWHM of 1.8--3.1~\AA, $<200$ $\rm km\ s^{-1}$; it is likely to be noise.)  \label{fig:Halpha_lim}}
\end{center}
\end{figure}

In \S \ref{sec:obs} we discuss our two reduction versions, ``A" and ``B," and Figure \ref{fig:comp5} shows that ``Redux A" has less noise in the region around H$\alpha$ owing to a more sophisticated background subtraction. With ``Redux A," we find that the maximum equivalent width of hydrogen allowed by the data is $W_{\lambda,A} (3\sigma) \approx 24.0$~\AA\ (for reduction version ``B" the result is $W_{\lambda,B} (3\sigma) \approx 42.9$~\AA, so our uncertainty here is $\Delta W_{\lambda,A} \approx 18$~\AA). This is shown in Figure \ref{fig:Halpha_lim}.

An emission line of peak $I=6.71\times10^{-17}$ $\rm erg\ s^{-1}\ cm^{-2}\ \AA^{-1}$ and $\rm{FWHM}=22$~\AA\ on top of our data would have a whopping equivalent width $W_{\lambda}(M_{\rm{H}}=0.05\ \rm{M_{\odot}}) \approx 4000$~\AA. If we scale linearly between the equivalent width and mass of hydrogen (as do Shappee et al. 2013a), then our upper limit on the amount of hydrogen is

\begin{equation}\label{eq:MH}
M_{\rm{H}} \lesssim \frac{24.0\ \rm{\AA}}{4000\ \rm{\AA}} \times 0.05\ \rm{M_{\odot}} \approx 3 \times 10^{-4}\ \rm{M_{\odot}},
\end{equation}

\noindent
with an uncertainty of $\Delta M_{\rm H} \approx 2.4 \times 10^{-4}\ \rm{M_{\odot}}$. Our results are very similar to the limit produced by Shappee et al. (2013a), $M_{\rm{H}} =5.8 \times 10^{-4}\ \rm{M_{\odot}}$, from a longer integration at an earlier phase. This uncertainty includes measurement errors only, and not systematics from model assumptions (see, e.g., Lundqvist et al. 2015). At this time, our estimate of $M_{\rm H}$ in Equation \ref{eq:MH} is essentially a lower limit on the upper limit of hydrogen in the system, and our main result is simply the qualitative nondetection of H$\alpha$.

\subsubsection{Broad H$\alpha$ at 6400~\AA?}\label{sssec:anaHbroad}

Taubenberger et al. (2014) attribute the new broad feature at 6400~\AA\ (seen in Figures \ref{fig:compspec} and \ref{fig:Halpha_lim}) to [\ion{Fe}{I}] or [\ion{O}{I}], and mention that detailed spectral modeling is required to confirm this. We point out that Mazzali et al. (2014) argue that the presence and strength of carbon at early times indicates there may be some hydrogen ($\sim0.01$ $\rm M_{\odot}$) in the outer layers of SN\,2011fe, which might be seen at high velocity (for SNe\,Ia, ``high velocity'' refers to 15,000--20,000 $\rm km\ s^{-1}$). Hydrogen at $\sim15,000$ $\rm km\ s^{-1}$ would appear at $\sim6250$~\AA, and given its location on the surface of the SN\,Ia --- not out in the CSM as considered in the analysis above --- may be seen as a broad emission line. If the broad feature we see at $\sim6400$~\AA\ is hydrogen, its velocity is $\sim7500$ $\rm km\ s^{-1}$, significantly lower than predicted. To estimate the equivalent width of the 6400~\AA\ feature in our 2014 spectrum, we aggressively bin (51 pixels, 32~\AA) and numerically integrate to find EW $\approx 700$~\AA. By Equation \ref{eq:MH}, this implies a mass of hydrogen of $M_{\rm{H}} \approx 0.009$ $\rm M_{\odot}$. Although this highly uncertain estimate is in good agreement with the predictions of Mazzali et al. (2014), the caveats described above regarding the conversion from equivalent width to mass apply here as well. Encouragingly, redshifted emission from hydrogen on the {\it far} side of the explosion may be seen at $\sim6700$~\AA\ in Figure \ref{fig:Halpha_lim}, but upon further inspection of the 2D images we find a diffuse extended artifact at that location, so this interpretation is not secure.

\subsection{Limits on the Luminosity and Temperature of a Post-Impact Remnant Star Companion}\label{ssec:anacomp}

In the single-degenerate scenario, when the companion star of an exploding white dwarf has its outer layers blown away by the supernova, the PIRS becomes hotter and more luminous. Models for nondegenerate He-burning companions (Pan et al. 2012, 2013) have shown they reach thousands of L$_{\odot}$ after 1--3~yr, and tens of thousands of L$_{\odot}$ after 10~yr. These PIRS can be modeled as blackbodies (BBs), and given their high temperatures they are significantly blue. At $\sim1$~yr after peak brightness, nebular emission from SN\,2011fe would swamp any contribution from a companion, but by $\sim3$~yr the contribution of a bright blue companion may be detectable. 

\begin{figure}
\begin{center}
\includegraphics[trim=1cm 0.2cm 0.0cm 0.5cm,clip=true,width=3.4in]{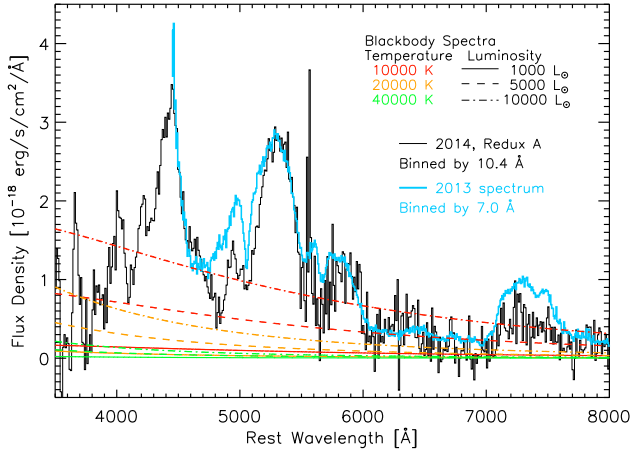}
\caption{The 2014 spectrum of SN\,2011fe (black) and the flux-scaled 2013 spectrum (blue; scaling described in \S~\ref{ssec:obscomp}) compared with blackbody spectra of temperatures and luminosities in the range predicted for a PIRS from Pan et al. (2012, 2013). \label{fig:compBB}}
\end{center}
\end{figure}

In order to constrain the contribution of a BB component in our 2014 spectrum of SN\,2011fe, we need a template $+981$ day nebular spectrum that is dominated by SN\,Ia flux to which we can add synthesised BB spectra. Since we do not yet have such a late-time spectrum for any other normal SN\,Ia, we instead use the 2013 spectrum, flux scaled to the 2014 spectrum as described in \S \ref{sec:obs}. (We do not use the synthesised 2012+2013 composite spectrum for this first method to constrain the BB component because it is sensitive to the flux calibration and changes in relative features, but we do use it below in Section \ref{sssec:anavar}). We then synthesise a set of BB spectra, and add them to the 2013 spectrum. To illustrate this idea, the 2013 spectrum is shown in Figure \ref{fig:compBB}, along with our 2014 spectrum of SN\,2011fe, and BB spectra with the range of temperatures and luminosities predicted for PIRS from Pan et al. (2013), to which our test is senstive: $\log (T_{\rm eff}/{\rm K}) = 4.0$--4.6 and $\log (L/{\rm L_{\odot}}) = 3.0$--4.0.

We compare this set of synthesised SN+BB spectra with the ``Redux A'' spectrum of SN\,2011fe from 2014 because it shows an excess of blue flux over ``Redux B" (\S \ref{sec:obs}), and will produce more conservative upper limits on the PIRS BB parameters. To statistically assess at what temperature and luminosity the SN+BB spectrum is significantly different from the 2014 spectrum of SN\,2011fe, we first flux-match the SN+BB spectrum to the 2014 spectrum on the red side (6000--7000~\AA). On the blue side, we then subtract the smoothed 2014 spectrum ($f_{\rm{SN}}$) from the SN+BB spectrum ($f_{\rm{SN+BB}}$) and divide this by the error spectrum of the 2014 data ($e_{\rm{SN}}$) to obtain a residual spectrum, 

\begin{equation} \label{eq:R}
\mathcal{R(\lambda)} = \frac{ f_{\rm{SN+BB}}(\lambda) - f_{\rm{SN}}(\lambda)  }{e_{\rm{SN}}}.
\end{equation}

\begin{figure*}
\begin{center}
\includegraphics[trim=0.0cm 0.2cm 0.0cm 0.5cm,clip=true,width=3.4in]{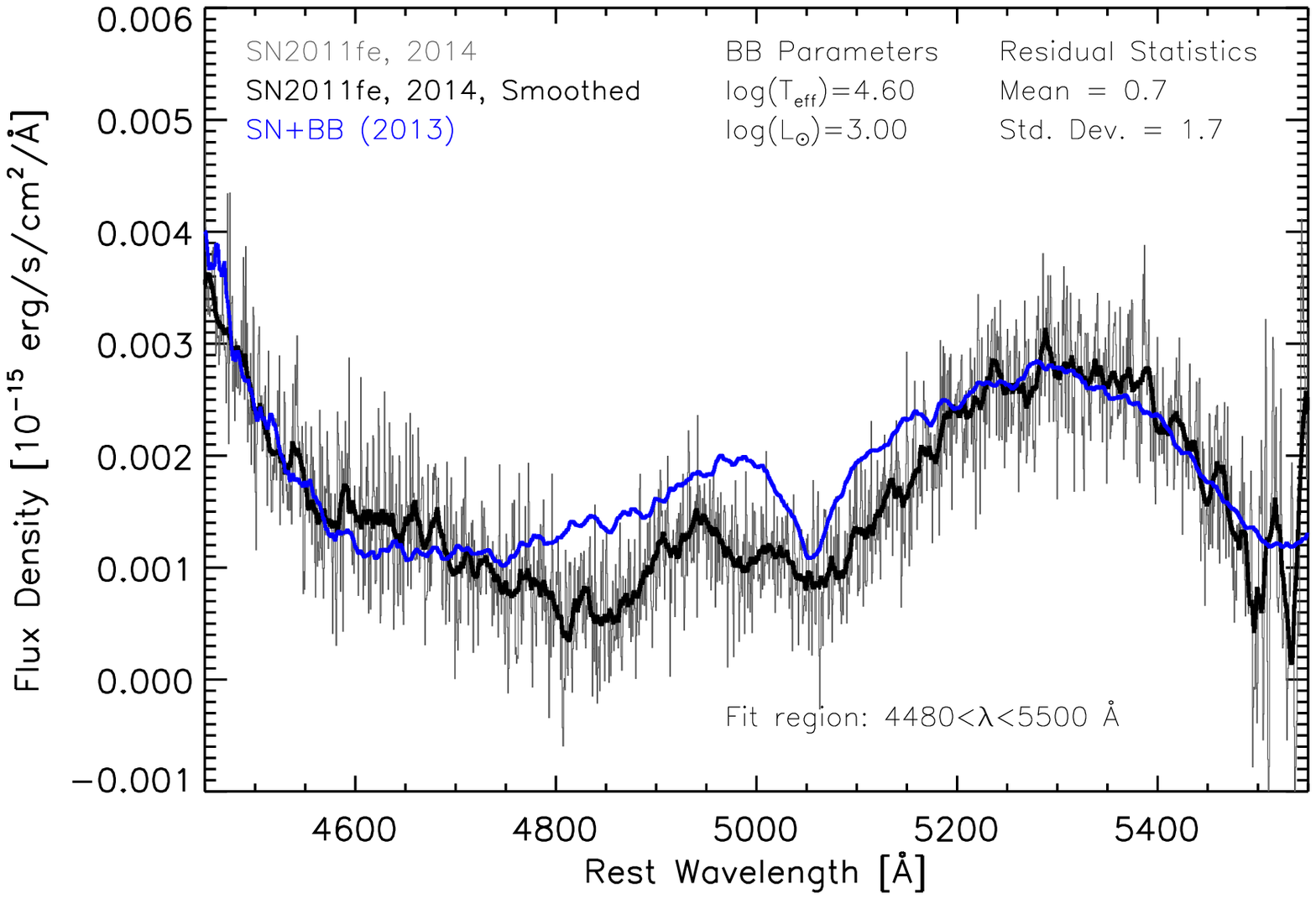}
\includegraphics[trim=0.0cm 0.2cm 0.0cm 0.5cm,clip=true,width=3.4in]{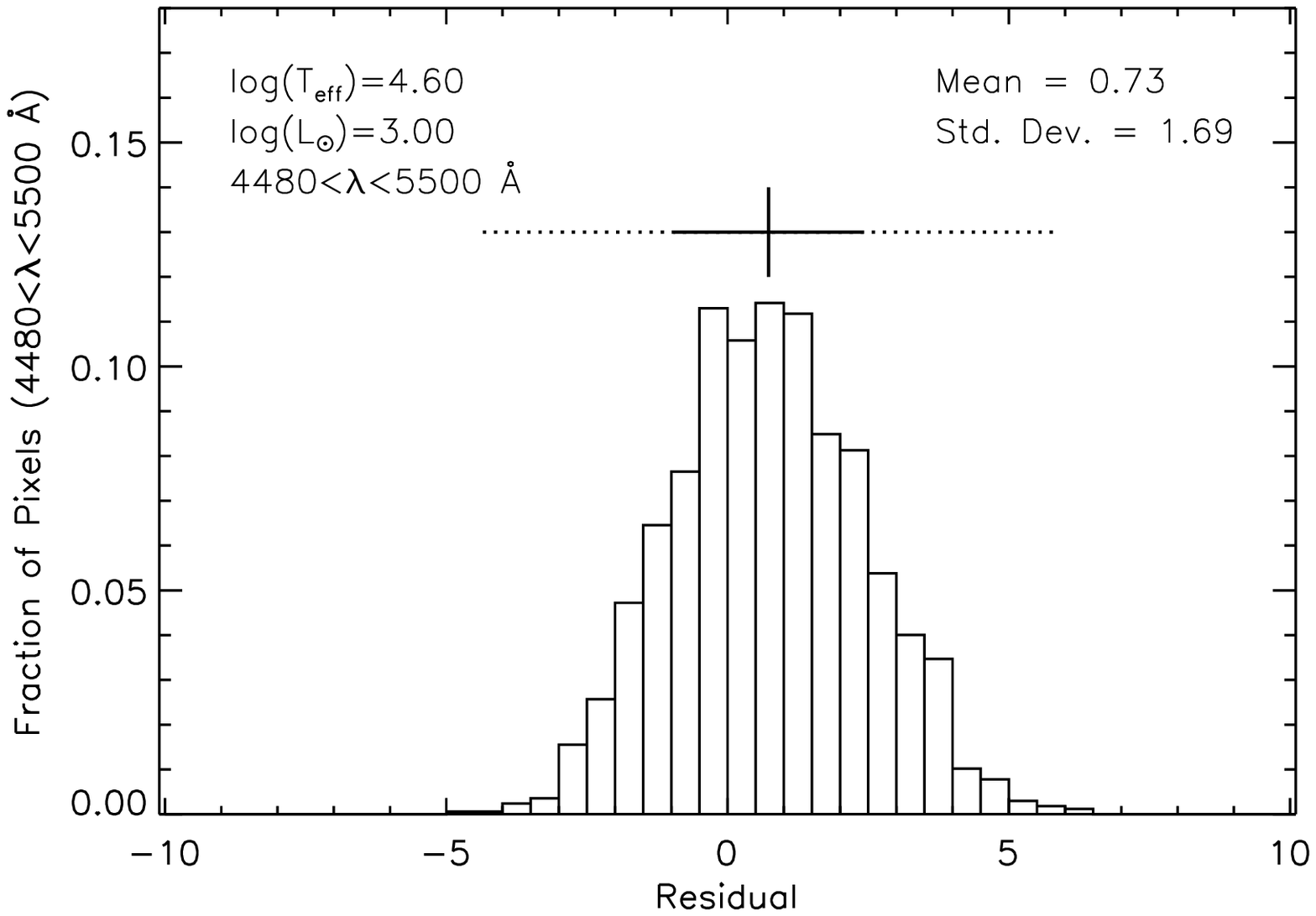}
\includegraphics[trim=0.0cm 0.2cm 0.0cm 0.5cm,clip=true,width=3.4in]{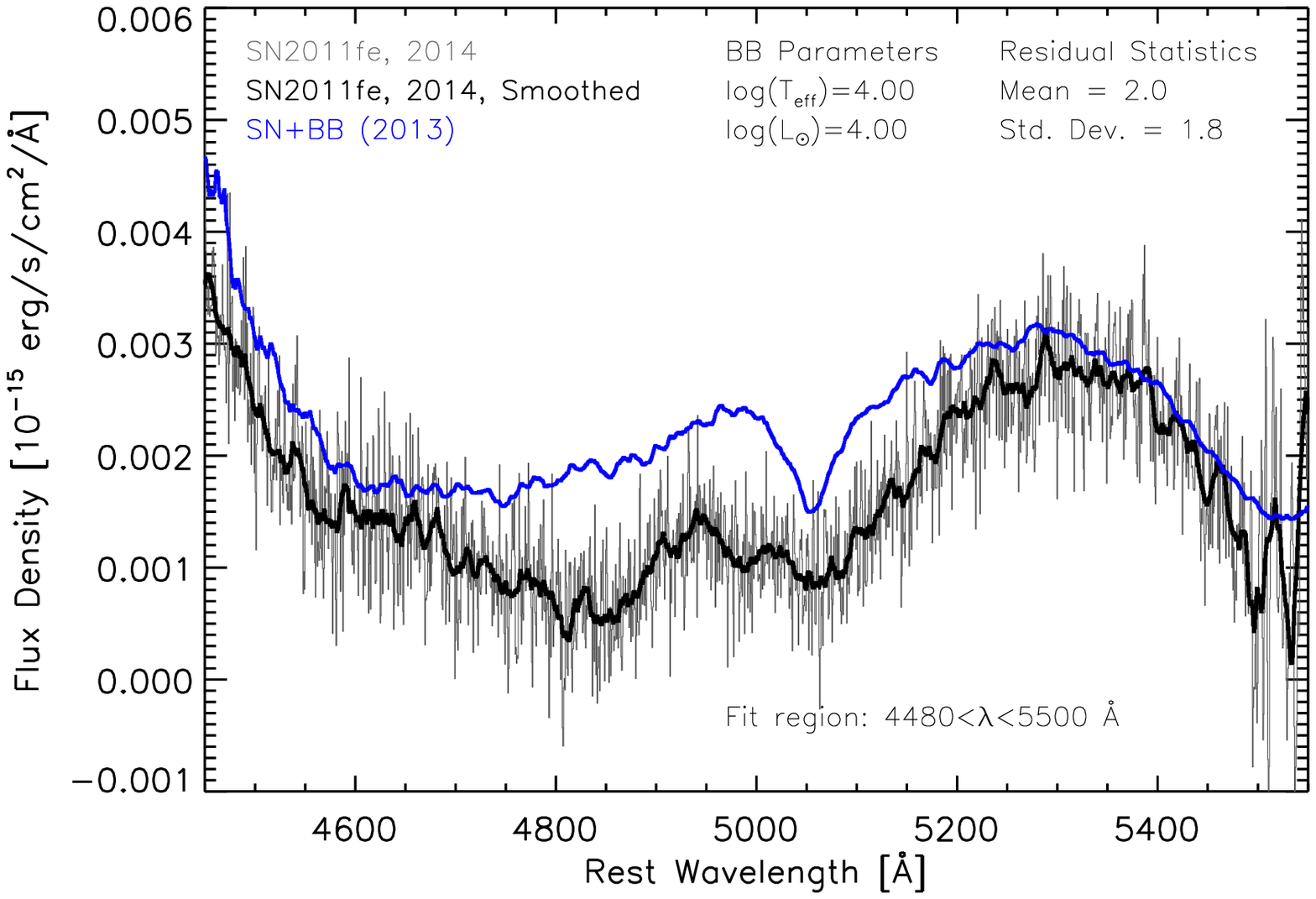}
\includegraphics[trim=0.0cm 0.2cm 0.0cm 0.5cm,clip=true,width=3.4in]{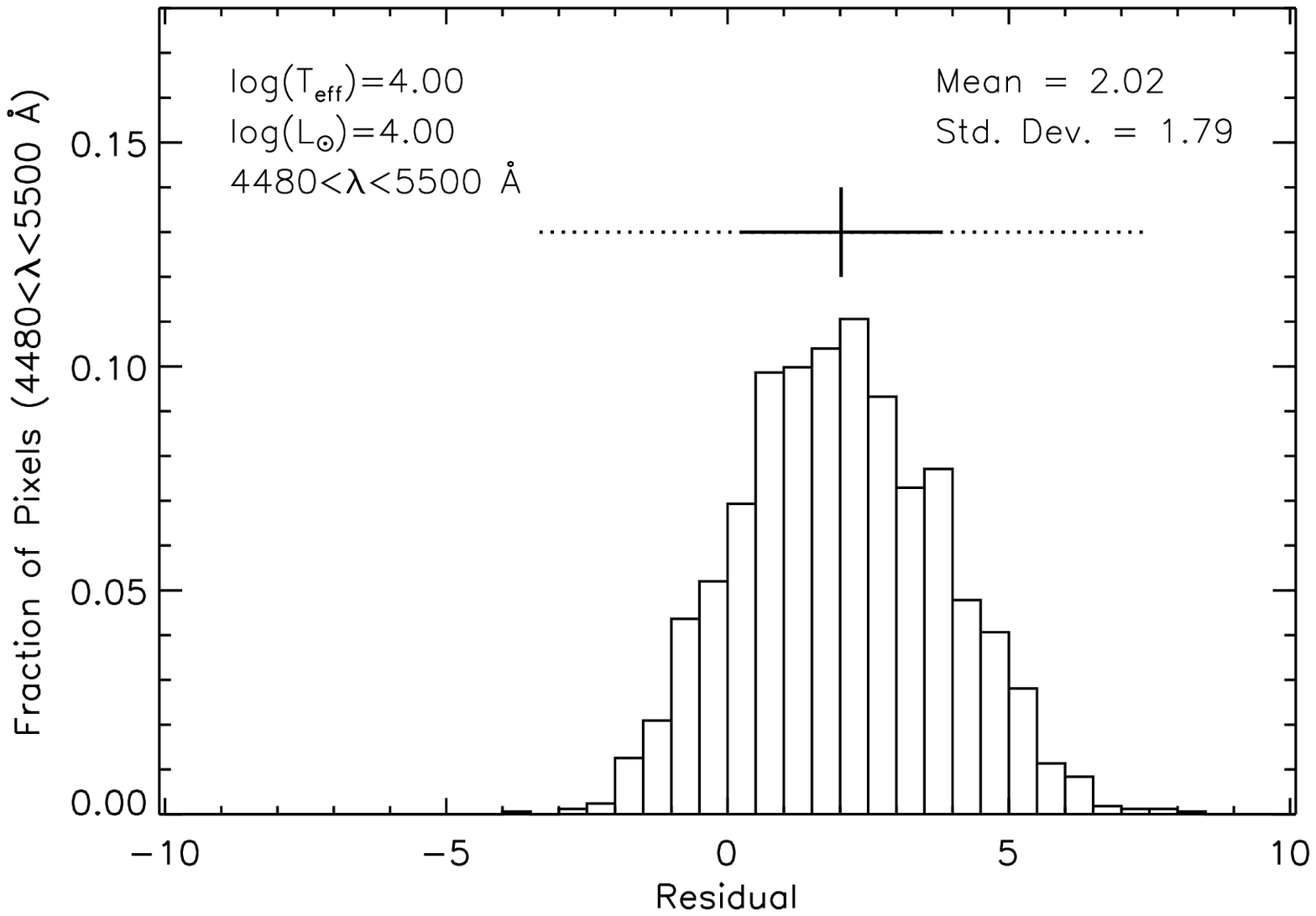}
\caption{\textbf{Left:} The original 2014 spectrum of SN\,2011fe (black, thin) and smoothed (black, thick), and the 2013 SN+BB spectrum (blue). Top and bottom rows show the two extrema models of nondegenerate PIRS companions as examples ($T_{\rm{eff}}$ and $L_{\odot}$ as in plot legend). \textbf{Right:} The distributions of residuals ($\mathcal{R}$ in Equation \ref{eq:R}), with the mean (vertical line), standard deviation (horizontal solid line), and three standard deviations (horizontal dotted line). \label{fig:compmodex}}
\end{center}
\end{figure*}

\noindent
In Figure \ref{fig:compmodex} we show examples of our SN+BB models compared with the 2014 spectrum of SN\,2011fe, and the resulting distribution of residuals, $\mathcal{R}$, at the blue end of the spectrum (4480--5500~\AA). When the mean and standard deviation of the residuals are inconsistent with zero, we can rule out the presence of a PIRS with the luminosity and temperature of the BB model added to our 2013 SN spectrum. This method is somewhat similar to the work of Fox et al. (2014), who search for the signature of a companion star in a spectrum of SN\,1993J. 

\begin{figure}
\begin{center}
\includegraphics[trim=1cm 0.2cm 0.0cm 0.5cm,clip=true,width=3.4in]{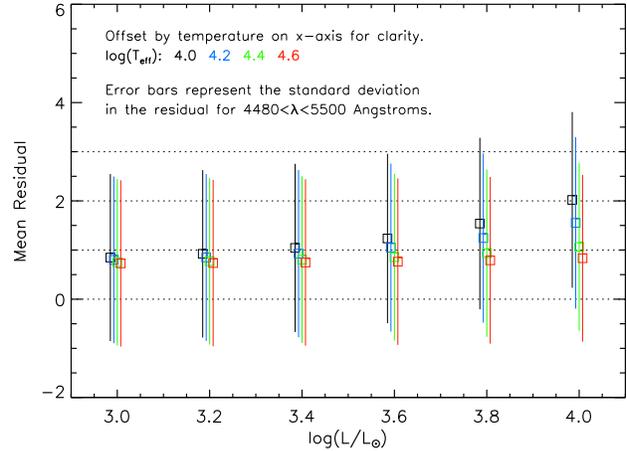}
\caption{The mean residual, $\bar{\mathcal{R}}$, in the 4480--5500~\AA\ range. Bars represent one standard deviation on the distribution of residual values (e.g., Figure \ref{fig:compmodex}). Horizontal dotted lines are drawn at $\bar{\mathcal{R}}$ = 0, 1, 2, and 3. It is evident how the temperature of the PIRS is not as constraining as the luminosity.  \label{fig:compmodfit}}
\end{center}
\end{figure}

In Figure \ref{fig:compmodfit} we plot the mean residual, $\bar{\mathcal{R}}$, for the variety of PIRS considered. We find that temperature is better constrained than luminosity, and that we can rule out the presence of a PIRS with  $T \lesssim 10000$ $\rm K$ and $L \gtrsim 10000$ $\rm L_{\odot}$ at the 2$\sigma$ level. As seen in Figure \ref{fig:compmodex}, within the spectral range of 4480--5500~\AA\ this test for a large change in the {\it continuum} is mostly insensitive to the small changes in the {\it features} between epochs -- but not entirely. The mean residual does not go to zero when the lowest-luminosity BBs are added because in the range of 4480--5500~\AA\ the 2014 spectrum's flux has declined with respect to the neighbouring broad feature at $\sim5400$~\AA, as seen in Figure \ref{fig:compBB}. Thus, our 2014 spectrum is sensitive only to the coolest and brightest BB components that can make an undeniably large change in the optical spectral continuum, over and above the existing small differences in features. 

Compared to the potential nondegenerate He PIRS models of Pan et al. (2013), our temperature and luminosity limits are not very constraining on the type of companion star at 3~yr. However, all model PIRS luminosity evolution tracks peak at $\sim10$~yr, and furthermore the full parameter space for all types of companions is not covered by these models; future theoretical work may find that other types of stars may be cooler and brighter. With patience we still have a chance, and there is some recent encouragement in this regard. Post-explosion imaging of SN\,Iax 2008ha obtained 4~yr after peak brightness shows a source, sufficiently nearby to be a PIRS from the progenitor system, that is $L \approx 10^4$ $\rm L_{\odot}$ (Foley et al. 2014). Future attempts to constrain the presence of a PIRS will have the benefit of our experience, and will be able to use this $\sim1000$ day spectrum of SN\,2011fe as their template.

\subsubsection{Variance Smoothing by a BB Component}\label{sssec:anavar}

\begin{figure}
\begin{center}
\includegraphics[trim=0.5cm 0.2cm 0.0cm 0.5cm,clip=true,width=3.4in]{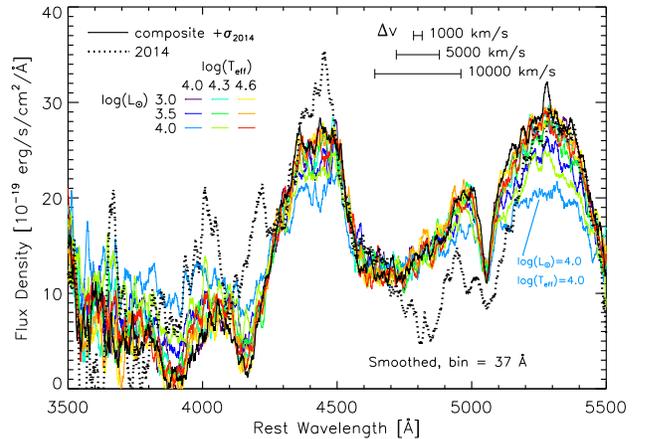}
\caption{To illustrate the effect of adding a BB component to a nebular SN\,Ia spectrum, we plot the 2014 spectrum of SN\,2011fe (black, dotted), the 2012+2013 composite spectrum (black, solid), and the composite SN+BB for a variety of BB temperatures and luminosities (coloured solid lines). All spectra are renormalised to have an integrated flux equal to the 2014 observations. The temperatures and luminosities of the BB spectra are given in the legend at left, and for the reader's reference we provide velocity scale bars in the top-right corner.  \label{fig:Vspec}}
\end{center}
\end{figure}

A second way to constrain the presence of a PIRS companion star in the 2014 spectrum of SN\,2011fe is to compare the amplitude of flux variation in 2014 with that of earlier spectra. A significant contribution from a blackbody spectrum will ``smooth out" the typical features of a nebular-phase SN\,Ia. For example, in Figure \ref{fig:Vspec} we compare the smoothed spectrum from 2014 with a variety of SN+BB spectra created using the 2012+2013 composite. These spectra have all been renormalised to have an integrated flux equal to that of the 2014 spectrum (i.e., they would produce equivalent photometric measurements). The amplitude of flux variation is clearly suppressed when we add BB components that are brighter at optical wavelengths.

\begin{figure}
\begin{center}
\includegraphics[trim=0.5cm 0.2cm 0.0cm 0.5cm,clip=true,width=3.4in]{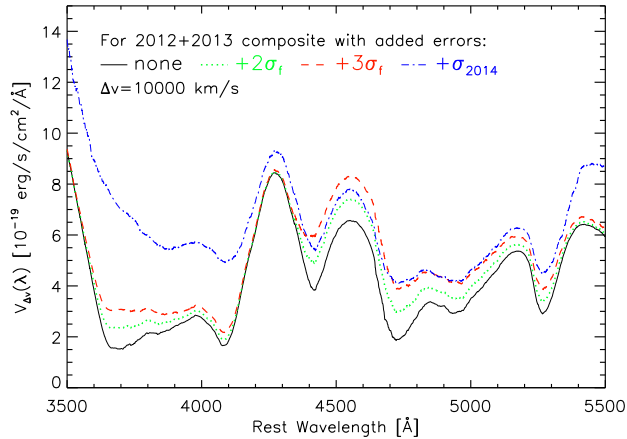}
\caption{Flux variance spectra, $V_{\Delta v}(\lambda)$, with a velocity scale of $\Delta v =10,000$ $\rm km\ s^{-1}$, for the SN\,2011fe composite 2012+2013 spectrum. To demonstrate the effect of flux-measurement uncertainty ($\sigma_f$) on the value of $V_{\Delta v}(\lambda)$, we add noise to the composite 2012+2013 spectrum and recalculate the variance. The black line represents the original measurement uncertainty; the green and red lines represent $2\sigma_f$ and $3\sigma_f$, respectively; and the blue line represents the addition of flux errors from the 2014 spectrum of SN\,2011fe, $\sigma_{2014}$. As expected, $V_{\Delta v}(\lambda)$ is higher for spectra with larger flux-measurement errors. \label{fig:Verr}}
\end{center}
\end{figure}

The flux variance, $V_{\Delta v}(\lambda)$, on a velocity scale of $\Delta v$ $\rm km\ s^{-1}$, for a given wavelength $\lambda$, is expressed by

\begin{equation}\label{eq:V}
V_{\Delta v}(\lambda) = \sqrt{ \frac{ \sum_{\lambda^{\prime}=\lambda_1}^{\lambda^{\prime}=\lambda_2} \left( f(\lambda^{\prime}) - \overline{f_{\Delta v}(\lambda)} \right)^2}{N_{\Delta v}(\lambda)}  },
\end{equation}

\noindent
where $\lambda_1$ and $\lambda_2$ are the wavelengths that correspond to  $\pm\Delta v$, $\overline{f_{\Delta v}}(\lambda)$ is the mean spectral flux within $\pm\Delta v$, the numerator is the sum of the residuals between the spectral flux and the mean spectral flux within $\pm\Delta v$, and the denominator is the number of pixels within $\pm\Delta v$. Although not exactly the statistical definition of variance, in its way $V_{\Delta v}(\lambda)$ is a measure of how much the spectrum varies in flux in a $2\Delta v$ $\rm km\ s^{-1}$ bin centred on $\lambda$. As an example, we show $V_{\Delta v}(\lambda)$ for $\Delta v = 10,000$ $\rm km\ s^{-1}$ for the 2012+2013 composite spectrum of SN\,2011fe in Figure \ref{fig:Verr}. 

The value of $V_{\Delta v}$ will also depend on the flux uncertainty, $\sigma_f$, or noise of the spectrum. To illustrate this, we take the flux error spectrum of the composite 2012+2013 spectrum, multiply it by a constant factor, randomise the sign, add it to the original composite spectrum, and recalculate $V_{\Delta v}(\lambda)$. In Figure \ref{fig:Verr}, the green and red lines show this for constant factors of 2 and 3, respectively, while the blue line shows this when the flux uncertainty of the 2014 spectrum is applied in the same fashion. The 2014 spectrum of SN\,2011 has the largest flux uncertainty.

We now use the flux variance to constrain the presence of a BB component in the 2014 spectrum of SN\,2011fe. The exact sequence of steps is to (1) compute the BB spectrum for a given temperature and luminosity and add it to the 2012+2013 composite spectrum for SN\,2011fe, (2) renormalise such that the SN+BB composite has the same integrated flux as the 2014 spectrum, (3) randomise the sign of the flux errors for the 2014 spectrum and add them to the renormalised SN+BB composite, and (4) compute the flux variance spectrum $V_{\Delta v}(\lambda)$. 

\begin{figure}
\begin{center}
\includegraphics[trim=0.5cm 0.2cm 0.0cm 0.5cm,clip=true,width=3.4in]{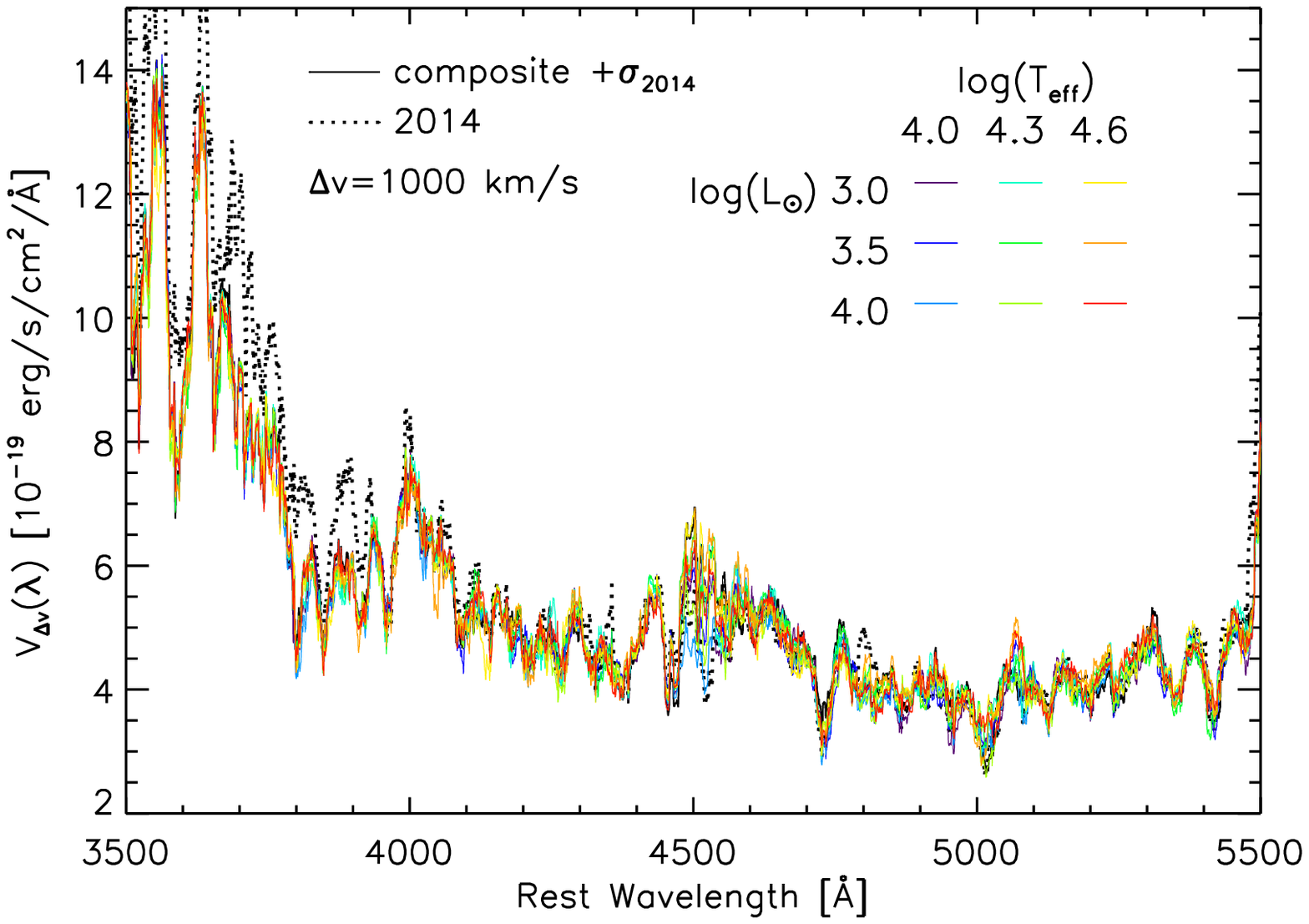}
\includegraphics[trim=0.5cm 0.2cm 0.0cm 0.5cm,clip=true,width=3.4in]{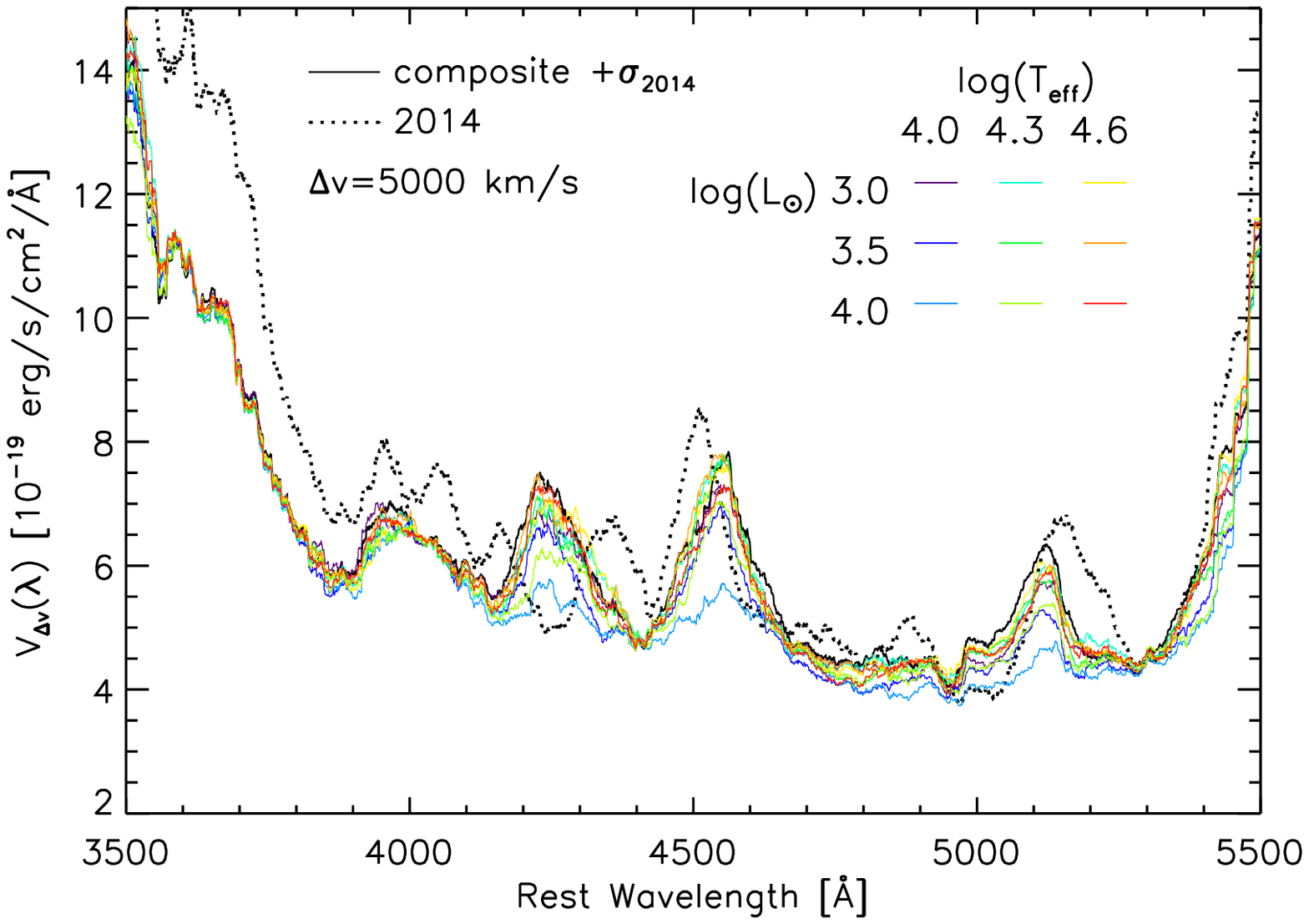}
\includegraphics[trim=0.5cm 0.2cm 0.0cm 0.5cm,clip=true,width=3.4in]{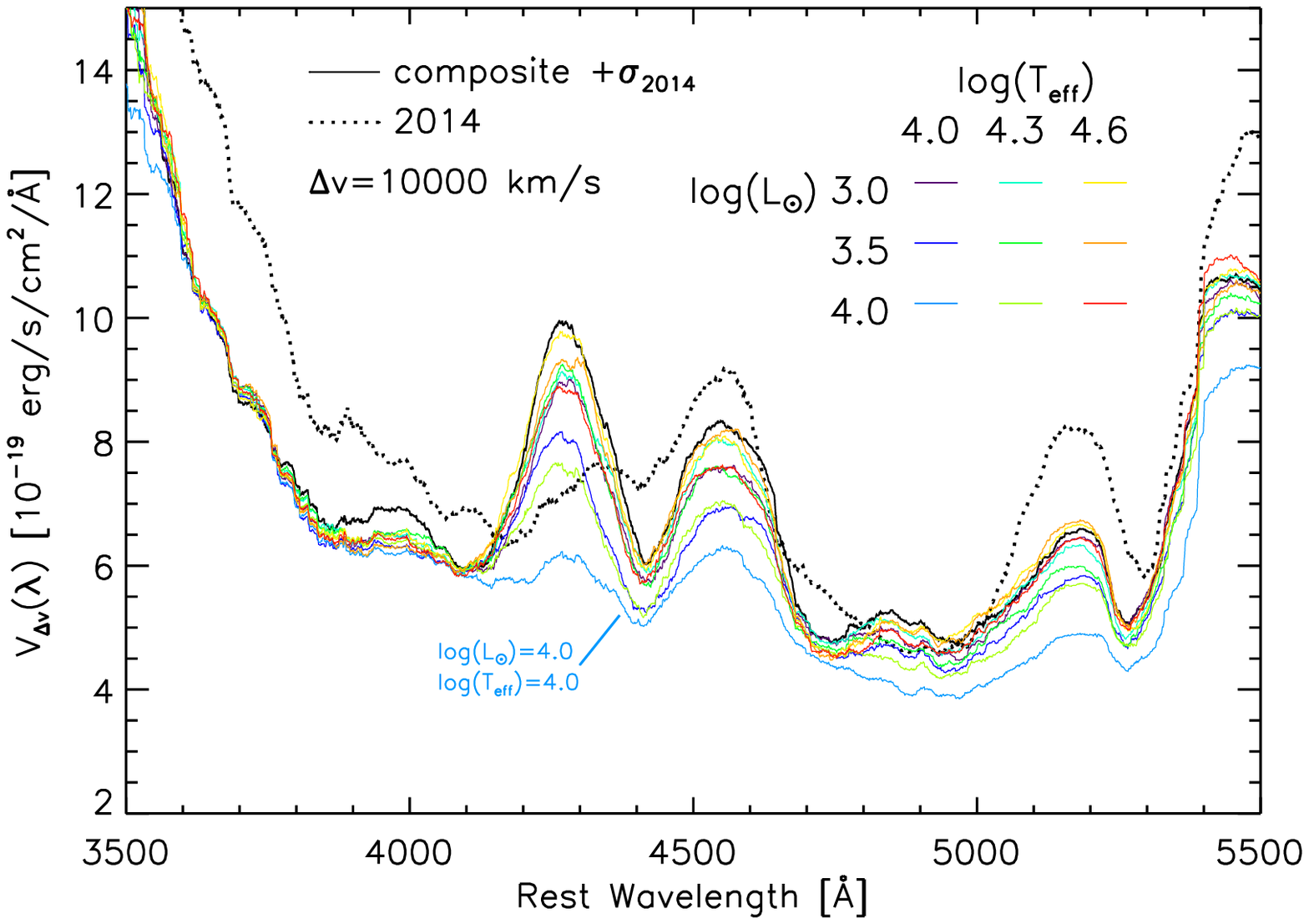}
\caption{Constraining the presence of a BB component in the 2014 spectrum of SN\,2011fe using the flux variance, $V_{\Delta v}(\lambda)$. From top to bottom, we show the flux variance when velocity scales of $\Delta v =$ 1000, 5000, and 10,000 $\rm km\ s^{-1}$ are used. In each plot, we show the variance for the 2014 spectrum of SN\,2011fe (black dotted line), the 2012+2013 composite spectrum with the flux uncertainties of 2014 applied (black solid line), and the composite SN+BB spectra (coloured solid lines, where colour represents BB temperature and luminosity as in the plot legend). Variance converges at the blue end where smaller-scale features and noise dominate.  \label{fig:V}}
\end{center}
\end{figure}

The typical velocity scale of features in SNe\,Ia is $\sim 10,000$ $\rm km\ s^{-1}$. In order to show the effect of $\Delta v$ on $V_{\Delta v}(\lambda)$, we include the results with 1000 and 5000 $\rm km\ s^{-1}$ in Figure \ref{fig:V} (for the size of $\Delta v$ in wavelength, see the top-right corner of Figure \ref{fig:Vspec}). A value $\Delta v= 1000$ $\rm km\ s^{-1}$ does not sample the nebular SN\,Ia features, only the noise, and in Figure \ref{fig:V} we can see that the effect of a BB component is not distinguishable ($\Delta v =1000$ $\rm km\ s^{-1}$ is included mainly as a sanity check). However, when we use $\Delta v= 5000$ or 10,000 $\rm km\ s^{-1}$, we find that the flux variance for the composite SN+BB spectrum looks distinctly different from the composite alone. While the variance spectra of the 2012+2013 composite and the 2014 spectrum are not exactly identical (solid and dotted black lines in Figure \ref{fig:V}), qualitatively they are more similar to each other than to a SN+BB with $T \sim 10000$ $\rm K$ and $\log (L/\rm{L_{\odot}}) \approx 4$. In this way, we judge that a bright, cool PIRS companion is unlikely for SN\,2011fe. Ultimately, this method of flux variance does not provide a tighter constraint on a putative PIRS than the flux-residual method described in \S \ref{ssec:anacomp}, but it is independent of any assumptions regarding evolution in the spectral features between the 2012, 2013, and 2014 spectra.

\section{Conclusions}\label{sec:con}

We demonstrate that extremely late-time photometry and spectroscopy of the nearby SN\,Ia 2011fe shows little evolution between $\sim2$ and $\sim3$~yr after explosion, and that there are several curious aspects of the nebular-phase features. There are new features at 5800 and 6400~\AA\ that could be attributed to [\ion{Fe}{I}], which may be consistent with the material cooling as indicated by the absence of  [\ion{Fe}{III}] and  [\ion{Co}{III}] at these late times. Alternatively, \ion{Na}{I}~D is a potential source for the 5800~\AA\ feature in a $\sim1000$ day spectrum. We also consider the possibility that the 6400~\AA\ feature is the broad, high-velocity hydrogen predicted by Mazzali et al. (2014). We find that a new emission feature has arisen between [\ion{Fe}{II}] $\lambda$7155 and [\ion{Ni}{II}] $\lambda$7370, which could be attributed to [\ion{Ca}{II}] $\lambda\lambda$7292, 7324. Emission from Na and Ca may suggest the presence of circumstellar material from the progenitor system. We show that the velocity of the nebular features has continued to evolve linearly at $\sim350$ $\rm km\ s^{-1}$ per 100 days, and discuss how the change in line morphology suggests that line blending may play a role, which argues against the hypothesis that nebular features are exhibiting a rest-frame redshift.

The presence of hydrogen in the CSM would support a nondegenerate companion for SN\,2011fe, but we do not see any evidence of narrow (FWHM $\approx$ 1000 $\rm km\ s^{-1}$) H$\alpha$ emission in our spectrum. Given this nondetection, we place an upper limit on the mass of hydrogen in the system, but we find that the models upon which this is based may not be appropriate for our extremely late-time spectrum. A lack of hydrogen in the progenitor system supports a degenerate companion for SN\,2011fe. This agrees with the pre-explosion archival $HST$ image analysis of SN\,2011fe presented by Li et al. (2011), which rules out red giants and most helium-star companions.

In this new, extremely late-time regime of SN\,Ia studies, the flux from the SN is sufficiently low that a contribution from a luminous secondary star could be detected. Although all preceding evidence indicates a degenerate companion for SN\,2011fe, in this paper we make the first observational tests for a nondegenerate PIRS. We synthesise a series of SN+BB spectra, and statistically constrain the temperature and luminosity for a putative companion with two independent techniques based on flux residuals and flux variance. We find that only the coolest, brightest PIRS models of Pan et al. (2013) are constrained by our observations. Although these tests are not currently sensitive enough to provide meaningful constraints on the progenitor scenario, the PIRS model's luminosity evolution does not peak until $\sim10$~yr after explosion --- so with patience we may yet improve in this regard.

\section*{Acknowledgements}

Based on observations from the Low Resolution Imaging Spectrometer at the Keck-1 telescope, the DEep Imaging Multi-Object Spectrograph at the Keck-2 telescope, and the Kast spectrograph on the 3-m Shane telescope. We thank the staff at the Lick and Keck Observatories for their assistance, Ori D. Fox for participating in the observations, and Peter Lundqvist, Josh Simon, and Ben Shappee for helpful correspondence. We thank Ken Shen for finding an error in our blackbody spectrum flux calibrations. Research at Lick Observatory is partially supported by Google. The W.~M.\ Keck Observatory is operated as a scientific partnership among the California Institute of Technology, the University of California, and NASA; it was made possible by the generous financial support of the W.~M.\ Keck Foundation. We wish to extend special thanks to those of Hawaiian ancestry on whose sacred mountain we are privileged to be guests. The supernova research of A.V.F.'s group at U.C. Berkeley is supported by Gary \& Cynthia Bengier, the Richard \& Rhoda Goldman Fund, the Christopher R. Redlich Fund, the TABASGO Foundation, and National Science Foundation (NSF) grant AST--1211916. M.S. acknowledges support from the Royal Society. J.M.S. is supported by an NSF Astronomy and Astrophysics Postdoctoral Fellowship under award AST-1302771.

\bibliographystyle{apj}
\bibliography{apj-jour,myrefs}

\label{lastpage}

\end{document}